hep-lat/9411006  6 Nov 94

# Flux tubes and topological charges in finite-temperature lattice QCD[†]

M. Faber[*], H. Markum[*], Š. Olejník[‡], W. Sakuler[*]

[*]*Institut für Kernphysik, Technische Universität Wien, Vienna, Austria*
[‡]*Institute of Physics, Slovak Academy of Sciences, Bratislava, Slovakia*

### Abstract

Topological properties of the vacuum were investigated in finite-temperature lattice QCD without and with dynamical quarks using a local operator of topological charge density and a variant of cooling. Below the deconfinement temperature evidence for flux tube formation was observed in the distribution of topological charge density around a static $Q\bar{Q}$ pair. With dynamical quarks, the flux tube breaks when the distance between the static quark and antiquark increases.

(hep-lat/yymmnnn)

# FLUX TUBES AND TOPOLOGICAL CHARGES
# IN FINITE-TEMPERATURE LATTICE QCD[†]

M. FABER, H. MARKUM, Š. OLEJNÍK* and W. SAKULER

*Institut für Kernphysik, Technische Universität Wien, A–1040 Vienna, Austria*

ABSTRACT

Topological properties of the vacuum were investigated in finite-temperature lattice QCD without and with dynamical quarks using a local operator of topological charge density and a variant of cooling. Below the deconfinement temperature evidence for flux tube formation was observed in the distribution of topological charge density around a static $Q\bar{Q}$ pair. With dynamical quarks, the flux tube breaks when the distance between the static quark and antiquark increases.

## 1. Introduction

Many phenomenological models of strong interactions motivated by QCD share a set of simple assumptions: confinement of quarks, antiquarks and gluons; nontriviality of the QCD ground state; existence of a flux tube between $Q$ and $\bar{Q}$ with constant energy density per unit length; existence of a phase transition temperature $T_c$ above which confinement and flux tubes disappear.

Lattice QCD is a powerful tool to test the above assumptions nonperturbatively from first principles. The existence of flux tubes has been proven in numerous lattice studies both indirectly, by showing that the $Q\bar{Q}$ potential rises linearly for large $Q\bar{Q}$ separations, and directly, by measuring correlations of plaquettes with the Wilson loop or a pair of Polyakov lines.[1,2]

Colour confinement is believed to be due to the structure of the QCD vacuum, which in turn should be a complicated superposition of quark and gauge fields with nontrivial topology. Then it is natural to expect that the existence of the flux tube between a $Q\bar{Q}$ pair will also be reflected in quantities characterizing topological properties of vacuum configurations containing a $Q\bar{Q}$ pair. The aim of the present contribution is to show that the existence and basic properties of flux tubes can be inferred from distributions of the topological charge density around a static $Q\bar{Q}$ pair.

## 2. Topological charge on a lattice and cooling

The topology of a continuum gauge field configuration is characterized globally by the (integer) value of the topological charge $\mathcal{Q}$:

---

[†]Presented by Š. Olejník. Work supported in part by BMWF.
*Permanent address: Institute of Physics, Slov. Acad. Sci., SK–842 28 Bratislava, Slovak Republic.

$$\mathcal{Q} = \int d^4x\, q(x) = \frac{g^2}{64\pi^2} \int d^4x\, \epsilon^{\mu\nu\rho\sigma}\, F^a_{\mu\nu}(x)\, F^a_{\rho\sigma}(x), \qquad (1)$$

where $q(x)$ is the topological charge density. A natural choice for defining $\mathcal{Q}$ on a lattice is to replace the integral in Eq. 1 by a sum over all lattice sites, $\mathcal{Q}_L = \sum_x q_L(x)$, with some lattice operator of topological charge density $q_L(x)$ which in the naive continuum limit converges to $q(x)$. Here we will restrict ourselves to results which we obtained using the *plaquette* definition[3]:

$$q_L(x) = -\frac{1}{2^4 32\pi^2} \sum_{\mu\nu\rho\sigma=\pm1}^{\pm 4} \tilde{\epsilon}_{\mu\nu\rho\sigma}\, \mathrm{Tr}\left[U_{\mu\nu}(x)\, U_{\rho\sigma}(x)\right], \qquad (2)$$

with $U_{\mu\nu}$ being a plaquette with $(\mu,\nu)$-orientation.* Employing this definition, a problem immediately arises: topological charges of lattice configurations are *non*-integer and in general rather small. The origin of this problem was clarified by Campostrini et al.[5]: lattice and continuum versions of the theory are different renormalized quantum field theories which differ from one another by finite renormalization factors, e. g. $q_L(x) = a^4\, Z(\beta)\, q(x) + O(a^6)$. $Z(\beta)$ is rather small for the range of $\beta$ values used in Monte Carlo simulations.

A way to get rid of renormalization constants while preserving physical information contained in lattice configurations is the *cooling* method.[6,7] One assumes that small local changes of a configuration do not change its global topological properties. We applied the variant of cooling suggested by Hoek et al.[7]

## 3. Results

Our results come from three runs, each of them performed on an $8^3 \times 4$ lattice using the Metropolis algorithm. In the first two runs we evaluated path integrals in pure SU(3) gauge theory with standard Wilson action for $\beta = 5.6$ (confinement phase) and 5.8 (deconfinement phase). The third run was done for QCD at $\beta = 5.2$ with SU(3) Wilson action for gluons and 3 flavours of Kogut–Susskind quarks of equal mass $ma = 0.1$ (confinement phase again). We made 100k iterations in each run and measured our observables after every 50th sweep. Each recorded configuration was subjected to 50 cooling steps.

In all simulations, cooling quickly removed the renormalization factor $Z(\beta)$ and the measured values of $\mathcal{Q}_L$ converged (close) to integer values. We then observed substantial differences between lattice configurations below and above the deconfinement phase transition. The $\mathcal{Q}_L = 0$ sector dominates in both phases, but in the confinement phase almost 70% of configurations have non-zero charge and the distribution of charges is approximately Gaussian (in both pure QCD and full QCD), while in the deconfinement phase $\mathcal{Q}_L \neq 0$ appears only very rarely.

The most interesting result comes from measuring $\langle L(0) L^\dagger(d)\, q_L^2(r)\rangle$, the correlation function which is related to (the square of) the topological charge density around

---

*Similar results were found with the *hypercube* definition of Di Vecchia et al.[3], see Ref. 4.

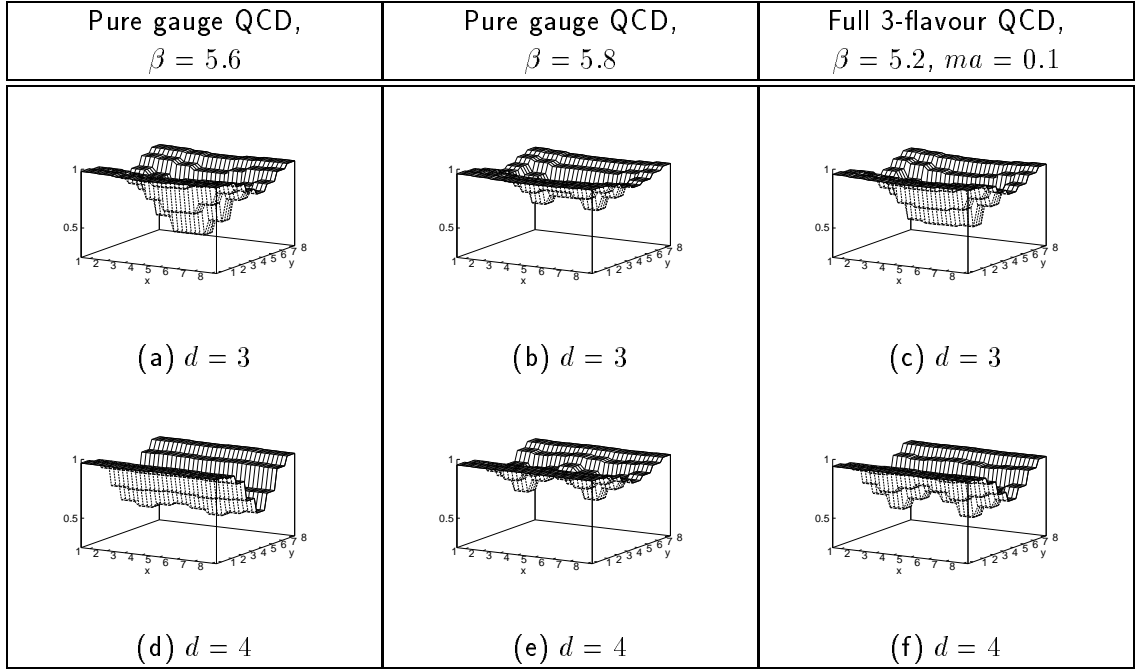

Fig. 1. The square of the topological charge density around a static quark-antiquark pair, $\langle L(0)L^\dagger(d)\, q^2(r)\rangle/(\langle L(0)L^\dagger(d)\rangle\langle q^2\rangle)$, at distances 3 and 4, after 5 cooling steps.

a static $Q\bar{Q}$ pair with separation $d$ ($L$ denotes the usual Polyakov line). In both phases we see a suppression of the topological charge density around static charges, consistent with phenomenological models. However, there is an essential difference between the phases which is illustrated in Fig. 1. In the deconfinement phase the suppression exists only in the vicinity of static sources (1b,e), while in the confinement phase it occurs in the whole flux tube between $Q$ and $\bar{Q}$, both without (1a,d) and with dynamical quarks (1c). In the pure gauge case the flux tube is most clearly visible at the largest separation, $d = 4$ (1d). In full QCD for $d = 4$ (1f) we see an indication of flux tube breaking due to creation of a virtual $q\bar{q}$ pair. A similar effect was observed earlier also in colour field distributions around a static $Q\bar{Q}$ system.[8]

## 4. References

1. For a review see A. Di Giacomo, *Acta Phys. Pol.* **B25** (1994) 215.
2. C. Schlichter, *this Conference*.
3. P. Di Vecchia et al., *Nucl. Phys.* **B192** (1981) 392.
4. M. Faber et al., *Phys. Lett.* **B334** (1994) 145.
5. M. Campostrini et al., *Phys. Lett.* **B212** (1988) 206.
6. M. Campostrini et al., *Nucl. Phys.* **B329** (1990) 683.
7. J. Hoek et al., *Nucl. Phys.* **B288** (1987) 589.
8. W. Feilmair and H. Markum, *Nucl. Phys.* **B370** (1992) 299.